\documentclass[a4paper,11pt,table]{article}
\usepackage{amssymb,amsmath,latexsym,color,amsthm,authblk,mathpazo,palatino,bbding,setspace,datetime,breakcites,hyphenat,microtype,diagbox,tikz,xcolor,subcaption,graphicx,titlesec}
\usepackage[T1]{fontenc}
\usepackage{newtxtext} 
\usepackage{mathptmx}
\usepackage[utf8]{inputenc}
\usepackage[round,authoryear]{natbib}
\usepackage{tikz-network}
\usepackage{hyperref}
\usepackage{theoremref}
\usepackage[titletoc,title]{appendix}
\usepackage{mathtools}
\usepackage{bm}
\usepackage{esvect}
\usepackage{mathtools}
\usepackage{bm}
\usepackage{esvect}
\usepackage{amsfonts}
\usepackage{mathrsfs}
\usepackage{amssymb}
\usepackage{physics}
\usepackage{yfonts}
\usepackage{dsfont}
\usepackage{multicol}
\usepackage{lipsum}
\usepackage{accents}
\usepackage{tikz-network}
\usepackage{csquotes}
\usepackage{titlecaps, stringstrings}
\usepackage{booktabs}
\usepackage{multirow}

\usepackage[shortlabels]{enumitem}
\usepackage{float}
\usepackage{amssymb}
\restylefloat{table}

\definecolor{armygreen}{rgb}{0.29, .8, 0.13}
\definecolor{auburn}{rgb}{0.43,0.21, 0.1}
\definecolor{burgundy}{rgb}{0.5,0.0, 0.13}
\definecolor{medium red}{rgb}{.490,.298,.337}
\definecolor{dark red}{rgb}{.235,.141,.161}

\hypersetup{
	colorlinks = true,
	linkcolor = {burgundy},
	citecolor = {burgundy},
	linkbordercolor = {white},
}

\let\OLDthebibliography\thebibliography
\renewcommand\thebibliography[1]{
	\OLDthebibliography{#1}
	\setlength{\parskip}{0pt}
	\setlength{\itemsep}{0pt plus 0.1ex}
}

\DeclareFontFamily{U}{mathx}{\hyphenchar\font45}
\DeclareFontShape{U}{mathx}{m}{n}{<-> mathx10}{}

\titleformat{\section}[block]{\normalfont\scshape\large\filcenter}{\thesection .}{1em}{}
\titleformat{\subsection}{\normalfont\scshape\large}{\thesubsection}{1em}{}
\titleformat{\subsubsection}{\normalfont\scshape\large}{\thesubsubsection}{1em}{}

\newtheorem{proposition}{Proposition}[section]

\newtheorem{lemma}{Lemma}[section]

\theoremstyle{definition}

\theoremstyle{remark}
\newtheorem{remark}{\textsc{Remark}}[section]

\newdateformat{monthyeardate}{\monthname[\THEMONTH], \THEYEAR}

\title{\textsc{}\thanks{\noindent}}

\author[1]{Agamani Saha\footnote{Contact: agamanisaha@gmail.com}}
\author[1]{Souvik Roy\footnote{Corresponding Author: souvik.2004@gmail.com}}
\affil[1]{Statistical Sciences Division, Indian Statistical Institute, Kolkata}
\date{}

\title{\textsc{Estimation of An Infinite Dimensional Transition Probability Matrix Using a Generalized Hierarchical Stick-Breaking Process}}
\begin{document}
	\maketitle
\begin{abstract}
Markov chains provide a foundational framework for modeling sequential stochastic processes, with the transition probability matrix characterizing the dynamics of state evolution. While classical estimation methods such as maximum likelihood and empirical Bayes approaches are effective in finite-state settings, they become inadequate in applications involving countably infinite or dynamically expanding state spaces, which frequently arise in domains such as natural language processing, population dynamics, and behavioral modeling. In this work, we introduce a novel Bayesian nonparametric framework for estimating infinite-dimensional transition probability matrices by employing a new class of priors, termed the Generalized Hierarchical Stick-Breaking prior. This prior extends traditional Dirichlet process and stick-breaking constructions, enabling highly flexible modelling of transition probability matrices. The proposed approach offers a principled methodology for inferring transition probabilities in settings characterized by sparsity, high dimensionality, and unobserved state spaces,  thereby contributing to the advancement of statistical inference for infinite-dimensional transition probability matrices.
\end{abstract}
\section{Introduction}
\subsection{Description and motivation of the problem} A Markov chain is a stochastic process that describes a sequence of possible events in which the probability of each event depends only on the state attained in the previous event. This property, known as the Markov property, simplifies the modeling of complex systems by focusing on immediate transitions between states rather than the full history of the process. A central component in the analysis of Markov chains is the transition probability matrix, which captures the likelihood of moving from one state to another in a single step. This matrix is square and stochastic, with each row summing to one, representing the transition probabilities from a given state to all possible states.

For finite-dimensional Markov chains, a variety of methods have been developed to estimate the transition probability matrix. One of the most widely used methods is the maximum likelihood estimator (MLE), which seeks the parameter values that maximize the likelihood of the observed transitions. The MLE for transition probabilities is intuitive and computationally straightforward: it is given by the relative frequencies of observed transitions between states. In addition to MLE, Bayesian and empirical Bayes approaches have gained increasing attention, particularly when dealing with sparse data or incorporating prior information.

While traditional estimation methods for finite-dimensional transition matrices such as maximum likelihood and empirical Bayes approaches have proven effective in a wide range of applications, they are often inadequate when dealing with systems that exhibit countably infinite state spaces. Many real-world processes, including natural language modeling, population dynamics, genetic sequence analysis, and customer behavior modeling, naturally give rise to Markov chains with an infinite or unbounded number of states. In such settings, the finite state assumptions become overly restrictive, and standard estimation techniques either break down.

\subsection{Contribution and novelty of the paper}
Despite the growing relevance of Markov chain models with a countably infinite number of state spaces, there remains a significant gap in the statistical literature concerning the estimation of transition probability matrices in such infinite-dimensional settings. To address this limitation, we introduce a novel Bayesian nonparametric framework for estimating transition probabilities, which is particularly well-suited for models with countably infinite number of state spaces. Specifically, we construct a model based on a generalized hierarchical extension of the traditional stick-breaking process, which serves as a highly flexible prior for the infinite-dimensional transition matrix, capable of accommodating a highly generalized correlation structure among the transition probabilities from the current state.

For posterior computation, we develop an efficient blocked Gibbs sampling algorithm that incorporates a set of carefully designed latent variables to facilitate conjugate updating and improve mixing. The proposed approach is empirically evaluated through simulations, where it is shown to outperform standard methods, such as the Maximum Likelihood Estimator, as well as the non-generalized version of the hierarchical stick-breaking prior in terms predictive accuracy. Notably, our method is capable of providing non-zero transition probability estimates for previously unobserved states in the data, and retains support over a theoretically infinite number of states. 

\subsection{Related Literature}
Bayesian nonparametric methods have emerged as a powerful alternative to classical parametric frameworks, especially in settings where the underlying structure is unknown or potentially infinite-dimensional. One of the most foundational developments in this area is the introduction of the Dirichlet Process (DP) by \citet{ferguson1973bayesian}, which serves as a prior distribution over probability measures.

A significant advancement in the usability of the Dirichlet Process came with the stick-breaking construction developed by \citet{sethuraman1994constructive}. This representation expresses a draw from the DP as an infinite sum of weighted atoms, where the weights are constructed via a stick-breaking process involving independent \(\text{Beta}(1, \beta)\) random variables. This formulation not only offers theoretical clarity but also enables practical computation through truncation approximations. The constructive nature of the stick-breaking process laid the groundwork for posterior inference algorithms and inspired further developments in the field.

Following this, \citet{ishwaran2001gibbs} extended the stick-breaking framework to develop Gibbs sampling techniques for a broad class of stick-breaking priors, offering a practical mechanism for performing Bayesian inference in models utilizing the DP. The approximation of DP measures via truncation was rigorously addressed by \citet{ishwaran2002exact}, who proposed both exact and approximate sum representations, enhancing computational efficiency without compromising theoretical integrity. Sampling strategies for key hyperparameters, particularly the concentration parameter \(\alpha\), were refined by \citet{liu2022sampling}, who developed efficient MCMC techniques to ensure stable and effective inference in high-dimensional applications.

The theoretical properties of these processes have been extensively studied. \citet{pitman2002poisson} introduced the Poisson-Dirichlet distribution and the associated GEM (Griffiths-Engen-McCloskey) distribution, both of which underlie the probabilistic behavior of the weights in the stick-breaking process. These distributions describe the law of the partitions induced by sampling from a DP and are central to understanding the asymptotic behaviour and clustering properties of nonparametric Bayesian models. In addition, generalizations of the Dirichlet distribution, such as those discussed by \citet{wong1998generalized} and \citet{connor1969concepts}, have enriched the landscape by accommodating more complex dependence structures among proportions.

The hierarchical extension of the Dirichlet Process, known as the Hierarchical Dirichlet Process (HDP), was introduced by \citet{teh2004sharing} to model grouped data where each group requires its own mixing distribution, yet with shared statistical strength across groups. This hierarchical modeling approach has proven highly effective in applications like topic modeling, speech processing, and biological sequence analysis. The nested HDP, introduced by \citet{paisley2014nested}, builds upon this by allowing deeper hierarchical structures, where the group-specific distributions themselves are organized within a tree-like hierarchy. Computational advances for HDPs include blocked Gibbs samplers proposed by \citet{das2024blocked}, which address the mixing issues associated with traditional samplers by enabling better exploration of the posterior space.

These developments in theory and computation have allowed Bayesian nonparametric methods to be applied to a wide range of practical problems. For example, \citet{beal2001infinite} introduced the infinite Hidden Markov Model (iHMM), which combines the HDP with Markovian dynamics to model sequences with an unknown number of latent states. The iHMM represents a nonparametric extension of the classical Hidden Markov Model (HMM), where the number of states is inferred from the data rather than fixed a priori. This has proven particularly useful in applications such as natural language processing, genomics, and behavioral modeling. The statistical underpinnings of such models are closely related to the Poisson-Dirichlet processes described earlier, which offer a natural basis for defining transition distributions in an infinite-state space.

The classical approach to estimating transition matrices in finite-state Markov chains is rooted in maximum likelihood estimation (MLE). Foundational work by \citet{anderson1957statistical} formalized inference in this context, demonstrating the asymptotic properties of MLEs and establishing their consistency and efficiency under certain ergodicity and stationarity assumptions. These methods rely on empirical transition counts and are straightforward when the number of states is small and the chain is well-behaved.

However, as the dimensionality of the state space grows, especially toward a countably infinite regime, these traditional techniques break down due to data sparsity and infinite parameter spaces. In response to this limitation, empirical Bayes methods have been proposed. Notably, \citet{seal2015empirical} developed an empirical Bayes framework for estimating finite-dimensional transition matrices. Their approach integrates prior distributions derived from the data itself and leverages numerical optimization and computational techniques, such as Gibbs sampling and EM-type algorithms, to improve estimation accuracy in small sample or sparse data regimes.

Despite their success in finite-state models, empirical Bayes and frequentist methods struggle to extend to infinite-dimensional contexts without strong truncation assumptions, which introduce bias and can degrade inferential quality. This shortcoming has motivated the use of Bayesian nonparametric methods to estimate infinite-dimensional transition matrices in a more principled way.

Recent literature has begun to address this challenge. For example, \citet{blei2010hdphmm} introduced the HDP-HMM, which places hierarchical Dirichlet process priors on the rows of the transition matrix, allowing for both infinite-state transitions and state sharing across sequences. While designed primarily for segmentation and clustering tasks, this model implicitly estimates a transition matrix with countably infinite rows. Similarly, the work of \citet{van2015markov} focuses on Markov exchangeable sequences, where the transition matrix itself is viewed as a random object drawn from a nonparametric prior, often a DP or Pitman-Yor process.

The theoretical underpinning for such models often draws from the structure of row-wise Dirichlet processes. Each row of the transition matrix is modeled as a draw from a DP (or its generalizations), ensuring that transitions from each state follow a discrete distribution over possibly infinite next states. \citet{james2009posterior} and \citet{lijoi2007bayesian} have contributed to the posterior analysis of normalized random measures, including Dirichlet and stable processes, which are often used to model rows of transition matrices in such frameworks.

Moreover, computational approaches have evolved to accommodate these infinite-dimensional objects. Blocked Gibbs samplers, slice sampling techniques (as in \citealt{walker2007sampling}), and truncation-free variational inference methods (e.g., \citealt{paisley2012stick}) have enabled tractable posterior inference in models with infinite transition matrices, supporting their application to real-world data.

Despite these advancements, relatively few works directly address the estimation of the transition probability matrix as a primary inferential target. The current literature is skewed toward latent variable inference, sequence labeling, and predictive modeling, often treating the transition matrix as a nuisance parameter. In contrast, many scientific and engineering applications, such as epidemic modeling, financial systems, and web user navigation, demand accurate, interpretable, and well-calibrated estimates of the full transition structure. The present work aims to bridge this gap by explicitly modeling the infinite-dimensional transition matrix using Bayesian nonparametric methods.

\section{Preliminaries}
We consider a discrete time Markov chain $\{X_1,X_2,\ldots \}$ with a countable state space $S=\{s_1,s_2,\ldots \}$. The transition probability matrix associated with this Markov chain is denoted by $\bm{P} = (P_{ij}: i,j \in S)$.
Let $x_1,x_2,\ldots,x_n$ be a (finite) realization of this process where $x_i \in S$ is an observed value of $X_i$. Let $n_{ij}$ be the number of times state $j$ is immediately preceded by state $i$ in the sample, that is, the process goes from state $i$ to state $j$ in one step. Our objective is to get an estimate of $\bm{P}$ based on the observed data $((n_{ij}))_{i,j=1}^d$. We consider a Bayesian setup where there is a prior belief about the distribution of the rows of the transition probability matrix $\bm{P}$.

\subsection{Dirichlet Distribution}

We define the strict $d-1$ dimensional simplex as   $\mathcal{S}_{d-1}:= \{ (x_1,\ldots,x_d) \in (0,1)^d \mbox{ such that } \sum_{i=1}^{d} x_i=1 \}$,  where $d \geq 2$. A Dirichlet distribution is a family of absolutely continuous multivariate probability distributions that is a multivariate generalization of the beta distribution and is used to model a vector of positive real quantities that sum to one. The random vector $\undertilde{X}=(X_1,X_2,\ldots,X_d)$ is said to follow a Dirichlet distribution of order $d \geq 2$ with parameters $\alpha_1, \alpha_2,\ldots,\alpha_d$, where $\alpha_i > 0$ for each $i$, that is $\undertilde{X} \sim$ $\text{Dir}(\alpha_1,\alpha_2,\ldots,\alpha_{d})$ if its probability density function can be expressed as
\[
    f(x_1,x_2,\ldots,x_d) = \frac{\Gamma(\alpha_0)}{\prod_{i=1}^d \Gamma (\alpha_i)} \prod_{i=1}^d x_i^{\alpha_i-1}
\]
where $\alpha_0 = \sum_{i=1}^d \alpha_i$ and $\{x_1,x_2,\ldots,x_d\}$ belongs to the strict $d-1$ simplex $\mathcal{S}_{d-1}$.

\subsubsection{Generalized Dirichlet Distribution}
In case of a Dirichlet distribution, the correlation between the $i$th and $j$th random variables where $i \neq j$ is given by $\frac{-\alpha_i \alpha_j}{\alpha_0+1}$, which implies that any pair of random variables are always negatively correlated. However, in practical scenarios, it may so happen that any pair of these variables may be positively correlated. Consider an example of a Markov chain which models the daily weather over a period of time. Suppose there are four types of weather - Sunny, Cloudy, Rainy and Windy. Now the correlation between the states Rainy and Cloudy is usually likely to be positive. In such a case Dirichlet distribution will not be an appropriate choice to model the distribution. To overcome this problem, we take recourse to a generalized version of the Dirichlet distribution. This distribution was developed by \cite{connor1969concepts}. It has almost twice the number of parameters as compared to the Dirichlet distribution but it has a more general covariance structure. This makes the generalized Dirichlet distribution a more practical and suitable choice to model this kind of data. The concise form of its probability density function was given by \cite{wong1998generalized}. According to this form, the random vector $\undertilde{X}=(X_1,X_2,\ldots,X_d)$ follows a Generalized Dirichlet Distribution with parameters $(\alpha_1,\alpha_2,\ldots,\alpha_{d};\beta_1,\beta_2,\ldots,\beta_{d})$, that is $\undertilde{X} \sim$ $GD(\alpha_1,\alpha_2,\ldots,\alpha_{d};\beta_1,\beta_2,\ldots,\beta_{d})$ if the probability density function of $\undertilde{X}$ can be expressed as
\[
    f(x_1,x_2,\ldots,x_{d}) = \prod_{i=1}^{d} \frac{1}{B(\alpha_i,\beta_i)} x_i^{\alpha_i-1}(1-x_1-x_2-\ldots-x_i)^{\gamma_i}
\]
where $\{x_1,x_2,\ldots,x_d,x_{d+1}\} \in \mathcal{S}^{d}$, and for all $i=1,2,\ldots,d$, $\gamma_i = \beta_i - \alpha_{i+1} - \beta_{i+1}$ for all $i=1,2,\dots,d-1$ and $\gamma_{d} = \beta_{d}-1$,  $ \alpha_i, \beta_i>0$, and $B(\alpha_i,\beta_i)$ is the Beta function given by $B(\alpha_i,\beta_i)=\Gamma(\alpha_i)\Gamma(\beta_i)/\Gamma(\alpha_i+\beta_i)$. 

The following lemma establishes the connection between Generalized Dirichlet Distribution and independent Beta random variables. 

\begin{lemma}[Construction of Generalized Dirichlet Distribution]\label{lemma0}
    Let $\undertilde{x} = (x_1,x_2,\ldots,x_d)$ be a random vector where each $x_j$ is independently and identically distributed as Beta$(\alpha_j,\beta_j)$. Consider the random vector $\undertilde{y} = (y_1,y_2,\ldots,y_d)$ such that
    \begin{align*}
        y_j & = x_{j} \prod_{k=1}^{j-1}(1-x_{k}) \qquad j=1,2,\ldots,d, 
    \intertext{Then,}
        (y_1,y_2,\ldots,y_{d}) & \sim \text{GD}(\alpha_1,\alpha_2,\ldots,\alpha_{d};\beta_1,\beta_2,\ldots,\beta_{d}). 
    \end{align*}
\end{lemma}
Proof is given in Appendix \ref{Appendix-B.1}.

\begin{remark}
    It follows from the definition that $
        \sum_{j=1}^{d}y_{j} < 1$.
\end{remark}

The moments of the Generalized Dirichlet Distribution have been provided by \cite{connor1969concepts}. Let $M_{i-1} = \prod_{k=1}^{i-1} \frac{\beta_k+1}{\alpha_k+\beta_k+1}$. Then, 
\begin{align*}
    E(X_i) & = \frac{\alpha_i}{\alpha_i+\beta_i} \prod_{k=1}^{i-1} \frac{\beta_k}{\alpha_k+\beta_k}, && i=1,\ldots,d\\
    V(X_i) & = E(X_i)\left[\frac{\alpha_i+1}{\alpha_i+\beta_i+1}M_{i-1} - E(X_i) \right]  && i=1,\ldots,d\\
    Cov(X_i,X_j) & = E(X_j) \left[\frac{\alpha_i}{\alpha_i+\beta_i+1}M_{i-1} - E(X_i) \right]  && i=1,\ldots,d-1;j=i+1,\ldots,d
\end{align*}
It is clear from the formula of the covariance that in this distribution, $X_1$ is always negatively correlated with the other random variables since $M_0 = 0$. However, $X_i$ and $X_j$ may be positively correlated for $i,j>1$. For example, if all the parameters of the distribution are equal, that is, $\alpha_i=\beta_i=\alpha$ for all $i=1,\ldots,d$, then it can be easily seen that $Cov(X_2,X_3)>0$ if $\alpha > 0$. Moreover, if there exists some $j>i$ such that $X_i$ and $X_j$ are positively (negatively) correlated, then $X_i$ will be positively (negatively) correlated with $X_k$ for all $k>i$. It is clear from the above formula that sign of the covariance term depends only on the first variable. Hence if $Cov(X_i,X_j)$ for any $j>i$, then the term inside the square brackets (involving $i$) in the covariance formula is positive (negative), which implies that $Cov(X_i,X_k)$ will be positive (negative) for all $k>i$.

It can be shown that the generalized Dirichlet distribution just like the Dirichlet distribution is conjugate to the multinomial distribution. 

\subsection{Dirichlet Process}
This section provides a brief overview of Dirichlet processes which can be thought of as an infinite generalization of the finite Dirichlet distribution. A Dirichlet process is a stochastic process that defines a distribution over probability distributions. It is used to generate random probability measures. Its existence was first proved by \cite{ferguson1973bayesian} in the context of developing a nonparametric prior for Bayesian inference. 

Let ($\Theta,\mathcal{B}$) be a measurable space where $\mathcal{B}$ be a Borel $\sigma$-field of subsets defined on $\Theta$. This $\mathcal{B}$ is the standard Borel $\sigma$-field which means there exists a topology $\mathcal{T}$ on $\Theta$ such that $(\Theta, \mathcal{T})$ is a (Polish) space and $\mathcal{B}$ is a Borel $\sigma-$algebra from that topology.

Let $\mathfrak{M} = \mathfrak{M}(\Theta)$ denote the set of probability measures on $(\Theta, \mathcal{B})$. It is equipped with the weak topology and the corresponding Borel $\sigma$-field $\mathcal{M}$.

Consider another probability space $(\Omega, \mathcal{A}, \mathbb{P})$. Let $G:(\Omega, \mathcal{A},\mathbb{P}) \rightarrow (\mathfrak{M}, \mathcal{M})$ be a measurable map. Then $G$ is called a random probability measure on $(\Theta, \mathcal{B})$. For any $\omega \in \Omega$, $G(\omega)$ is an element of $\mathfrak{M}$, that is a probability measure on $(\Theta, \mathcal{B})$. For any measurable set $A \in \mathcal{B}$, the evaluation map $e_A:\mathfrak{M} \rightarrow [0,1]$ defined by $e_A(\mu) = \mu(A)$ is measurable, and the composition $e_A \circ G: \Omega \rightarrow [0,1]$ is a real-valued random variable, denoted by $G(A)$, which is $\mathbb{P}$-measurable. 

Let $G_0$ be a fixed probability measure defined on the measurable space $(\Theta, \mathcal{B})$. Let $\alpha_0$ be a positive real number. Now the random probability measure $G$ is said to follow a Dirichlet process $DP(\alpha_0,G_0)$ with base measure $G_0$ and concentration parameter $\alpha_0$ if for any finite measurable partition $(A_1,A_2,\ldots,A_r)$ of $\Theta$, the random vector $(G(A_1),\ldots,G(A_r))$ is distributed as a Dirichlet distribution with parameters $(\alpha_0 G_0(A_1),\ldots,\alpha_0 G_0(A_r))$, that is,
\begin{align}
     (G(A_1),\ldots,G(A_r)) \sim \text{Dirichlet}(\alpha_0 G_0(A_1),\ldots,\alpha_0 G_0(A_r)) 
     \label{eq0}
\end{align} 
We write $G \sim DP(\alpha_0,G_0)$. The Dirichlet process is a probability measure on $(\mathfrak{M}, \mathcal{M})$ whose realizations are probability measures on $(\Theta, \mathcal{B})$.

\subsubsection{Hierarchical Dirichlet Process}
Now we move on to the problem where the data is divided into a countable number of groups. In this context the Hierarchical Dirichlet Process (HDP) (\cite{beal2001infinite}, \cite{teh2004sharing}) is used which is an extension of the previously described Dirichlet Process. It introduces a hierarchical structure, allowing multiple related groups to share common components, while allowing each group to have its own specific distribution over these components. HDPs are particularly useful for modeling data that can be grouped into related subsets, where each subset can share some latent structure (like clusters or topics), while simultaneously permitting for variation among the groups. In our problem, we consider each row of the transition probability matrix as a group. The  groups (rows) are indexed as $i=1,2,\dots$. 

Consider a collection of random probability measures $\{G_i:(\Omega, \mathcal{A}, \mathbb{P}) \rightarrow (\mathfrak{M}, \mathcal{M})\}_{i \in \mathcal{I}}$, where $\mathcal{I}$ is the index set of groups, and a global random random probability measure $G:(\Omega, \mathcal{A}, \mathbb{P}) \rightarrow (\mathfrak{M}, \mathcal{M})$. The global measure $G$ follows a Dirichlet process with concentration parameter $\alpha_0$ and a base (fixed) probability measure $G_0$ on $(\Theta, \mathcal{B})$.
\[
    G \sim \text{DP}(\alpha_0,G_0)
\]

The random measures $G_i$ are conditionally independent given $G$, with distributions given by a Dirichlet process with base probability measure $G$ such that
\begin{align}
    G_i|G \sim \text{DP}(\beta_0,G)
\end{align}
    
For each $\omega \in \Omega$, define $supp(\mu(\omega))$ as the support of the deterministic probability measure $\mu(\omega)$. For each outcome $\omega$, the support $supp(\mu(\omega))$ is a specific subset of $\Theta$. Now for HDP, we have for almost every $\omega \in \Omega$
\[
    supp(G_i|G(\omega)) \subseteq supp(G(\omega))
\]

The hyperparameters of the HDP are the baseline probability measure $G$, and the concentration parameters $\alpha_0$ and $\beta_0$. If the variability of different groups are expected to be different, a separate concentration parameter $\beta_{0i}$ may be used for each group. In this paper we consider $\beta_{0i}=\beta_0$ for all $i$. 

\section{Stick-Breaking Process}
The concept of the stick-breaking process involves breaking a stick of unit length into pieces repeatedly to generate a sequence of proportions. In the literature, it has been defined as follows.
Suppose
\[
    \nu_{j} \overset{i.i.d}{\sim} \text{Beta}(1, \beta) \qquad j=1,2,\ldots
\]
Let $\undertilde{\gamma}=\{\gamma_1,\gamma_2,\dots\}$ be such that $\gamma_j:\Omega \rightarrow [0,1]$ and $\undertilde{\gamma}$ lies in the infinite dimensional simplex $\mathcal{S}_{\infty} = \{(\gamma_1,\gamma_2, \dots )\in (0,1)^\mathbb{N}:\sum_{j=1}^{\infty}\gamma_j=1)\}$. Then the process $\{\gamma_1,\gamma_2,\dots\}$ is said to follow a stick breaking process if
\[
    \gamma_{j} = \nu_{j} \prod_{k=1}^{j-1}(1-\nu_{k}) \qquad j=1,2,\ldots
\] 

It was known in the literature much long ago as the Residual Allocation Model (RAM) or as the model with $\nu_1,\nu_2,\ldots$ as (discrete) failure rates. The distribution of the random discrete distribution $\undertilde{\gamma} = (\gamma_1, \gamma_2,\ldots)$ is also known as the GEM$(\beta)$ or GEM$(\undertilde{\gamma})$
(Griffith-Engen-McCloskey) distribution. The details of this distribution was discussed by \cite{pitman2002poisson}. This process was used in the context of Dirichlet process by \cite{sethuraman1994constructive}, which is popularly known as the Sethuraman construction of Dirichlet priors. Accroding to this construction, the stick-breaking process samples from the Dirichlet process, where each break in the stick corresponds to a component's probability mass, and the process of generating these breaks follows the Dirichlet process's definition. Specifically, if $G \sim \text{DP}(\alpha,G_0)$, then the Stick-Breaking Process generates the random probability measure $G$ as
\[
    G = \sum_{j=1}^\infty \gamma_j \delta_{\theta_j}
\]
where $\gamma_j$ is as defined before according to the Stick-Breaking Process and $\delta_{\theta_j}$ is the Dirac delta measure at the point $\theta_j:\Omega \rightarrow \Theta$ which is defined as a map
\begin{align*}
    \delta_{\theta_j} & : \Omega \rightarrow \{0,1\}\\
    \delta_{\theta_j}(A) & = \begin{cases}
        1 \text{, if } \theta_j \in A \text{ for any measurable set } A \subseteq \mathcal{B}\\
        0 \text{, o.w.}
    \end{cases}
\end{align*}
$\theta_j$'s are independent and identically distributed draws from the base distribution $G_0$ and they are independent of $\gamma_j$'s.

\subsection{Generalized Stick-Breaking Process} \label{GSBP}
In the traditional Stick-Breaking Process as described in the previous section the first shape parameter of the beta distribution was fixed at 1. This restricts the shape of the distribution to either increasing, decreasing, or a ``U"-shaped function. If the distribution is of any other shape, say, inverted "U" shape, that is, the mid-values have highest probability of occurrence, then the above Stick-Breaking process will not be appropriate. To overcome this drawback, we generalize the Stick-Breaking Process as follows.
Suppose
\begin{align}
    \nu_{j} \overset{i.i.d}{\sim} \text{Beta}(\alpha, \beta) \qquad j=1,2,\ldots
\end{align}
As in the previous section, let $\undertilde{\gamma}=\{\gamma_1,\gamma_2,\dots\}$ be such that $\gamma_j:\Omega \rightarrow [0,1]$ and $\undertilde{\gamma}$ lies in the infinite dimensional simplex $\mathcal{S}_{\infty} = \{(\gamma_1,\gamma_2, \dots )\in (0,1)^\mathbb{N}:\sum_{j=1}^{\infty}\gamma_j=1)\}$. Then the process $\{\gamma_1,\gamma_2,\dots\}$
is said to follow a Generalized Stick-Breaking Process where $\sum_{j=1}^{\infty}\gamma_{j}=1$ if
\begin{align}
    \gamma_{j} = \nu_{j} \prod_{k=1}^{j-1}(1-\nu_{k}) \qquad j=1,2,\ldots
\end{align}
We have named this process as GGEM$(\alpha,\beta)$. This is a generalized version of the GEM($\alpha$) distribution. GGEM stands for Generalized Griffiths, Engen and McCloskey Distribution.
\subsection{Hierarchical Generalized Stick-Breaking Process}
In a hierarchical Stick-Breaking Process (\cite{paisley2014nested}), a stick is broken into 'super-components' using a Dirichlet process. Each 'super-component' is then further decomposed using another stick-breaking process to create 'sub-components'. For our purpose, we consider a Hierarchical Generalized Stick-Breaking Process that introduces a hierarchy in the Generalized Stick-Breaking Process as defined in Section \ref{GSBP}. Mathematically, it can be represented as follows. Let $\gamma_j$ be as defined beforehand, that is,
    \begin{align}
        \gamma_j & = \nu_{j} \prod_{k=1}^{j-1}(1-\nu_{k}) \qquad j=1,2,\ldots \text{ and }
        \sum_{j=1}^{\infty}\gamma_{j}=1
    \intertext{where,}
        \nu_{j} & \overset{i.i.d}{\sim} \text{Beta}(\alpha,\beta) \qquad j=1,2,\ldots
    \end{align}
Let $(\Omega, \mathcal{A}, \mathbb{P})$ be a probability space as defined above. Now, let $\Theta = \mathbb{N}$, the set of all natural numbers, and, $\mathcal{B}=\mathit{P(\mathbb{N})}$ be the power set of $\mathbb{N}$. Let $\mathfrak{M}$ be the set of all proabability measures on $(\mathbb{N},\mathit{P(\mathbb{N})}$ with corresponding Borel $\sigma$-field $\mathcal{M}$. Define a collection of random probability measures $\{\undertilde{\pi_i}:(\Omega,\mathcal{A},\mathbb{P}) \rightarrow (\mathbb{N}, \mathit{P(\mathbb{N})})\}_{i \in \mathbb{N}}$, where $\pi_i= (\pi_{ij})_{j=1}^{\infty} = (\pi_{i1},\pi_{i2},\dots)$. For any measurable set $A \subseteq \mathit{\mathbb{N}}$, the evaluation map $e_A:\mathfrak{M} \rightarrow [0,1]$ is defined by $e_A(\undertilde{\pi}) = \undertilde{\pi}(A) = \sum_{k \in A} \pi_k$. Now let us assume,
    \begin{align}
        \undertilde{\pi_i}|\undertilde{\gamma} & \sim \text{DP}(\alpha_0, \undertilde{\gamma}), \qquad \text{ for all } i=1,2,\dots \label{eq6}
    \end{align}
This implies that for any finite measurable partition $(A_1,A_2,\dots,A_r)$ of $\mathbb{N}$, we have,
\[
 \Big(\undertilde{\pi}(A_1),\dots,\undertilde{\pi}(A_r)\Big) \Big{|} \undertilde{\gamma} \sim \text{ Dirichlet} \Big(\alpha_0\undertilde{\gamma}(A_1),\dots \alpha_0 \undertilde{\gamma}(A_r) \Big)
\]
\begin{lemma} \label{lemma1}
    If a random vector $\undertilde{X}=(X_1,X_2,\ldots,X_k)$ follows a Dirichlet distribution with parameters $(\alpha_1,\alpha_2,\ldots,\alpha_k)$, then
     \begin{align}
        \frac{1}{1-X_1}(X_2,X_3,\ldots,X_k)  \sim \text{Dirichlet}(\alpha_2,\alpha_3,\ldots,\alpha_k)
    \end{align}
Proof is provided in Appendix \ref{Appendix-B.2}.
\end{lemma}
    \begin{proposition} \label{GSBP_proof} If the random vector corresponding to the $i$th row of the TPM follows a Dirichlet Process, that is,
        \begin{align}
            \undertilde{\pi_i} | \undertilde{\gamma} & \sim \text{DP}(\alpha_0, \undertilde{\gamma})\\
        \intertext{Then it implies that}
            \pi_{ij} & = \pi'_{ij}\prod_{k=1}^{j-1}(1-\pi'_{ik}) \label{eq2}
        \intertext{where,}
            \pi'_{ij}|(\gamma_1,\gamma_2,\dots,\gamma_j) & {\sim}\text{ Beta }(\alpha_0\gamma_j, \alpha_0(1-\sum_{k=1}^{j}\gamma_k)) \qquad i,j=1,2,\ldots \label{eq1}
        \end{align} 
    \end{proposition}
Proof is provided in Appendix \ref{Appendix-B.3}.
\section{The Model}
We have a Markov chain $\{X_1,X_2,\ldots.\}$ with a countable state space $S=\{s_1,s_2,\ldots\}$. The transition probability matrix associated with this Markov chain is denoted by $\bm{P} = (P_{ij}: i,j \in S)$ with $P_{ij} \geq 0$ and $\sum_{j} P_{ij} = 1$ for all $i,j \in S$.
Let $x_1,x_2,\ldots,x_n$ be the realization of this process where $x_i \in S$, which has been observed. Note that we have observed only a finite number of states in our sample, while the number of states is actually countably infinite. Let $N_{ij}$ denote the count random variable of moving from state $i$ to state $j$. Let $n_{ij}$ be the number of times state $j$ is immediately preceded by state $i$ in the sample, that is, the process goes from state $i$ to state $j$ in one step. Let $\undertilde{n_i} = (n_{i1},n_{i2},\ldots,n_{ij},\ldots)$. Let $\pi_{ij}$ denote the true transition probability of moving from state $i$ to state $j$. Let $F(\undertilde{\pi_i})$ denote the distribution of $\undertilde{N_i}$ given $\undertilde{\pi_i}$. We put a Hierarchical Generalized Stick-Breaking Prior on $\undertilde{\pi_i}$. For each i, the model is given as $\undertilde{\pi_i}|G_i \sim G_i$, where $G_i$ denotes the prior distribution for $\undertilde{\pi_i}$. Let DP$(\alpha,H)$ denote a Dirichlet process with base measure $H$ and concentration parameter $\alpha$. The Generalized Hierarchical Stick-Breaking Process defines a global Generalized Stick-Breaking Process GGEM$(\alpha,\beta)$ such that $\undertilde{\gamma} \sim$ GGEM$(\alpha,\beta)$ and a set of random probability measures $G_i \sim$ DP$(\alpha_0,\undertilde{\gamma})$ which are conditionally independent given $\undertilde{\gamma}$. Proposition $\ref{GSBP_proof}$ implies that this is equivalent to $\pi_{ij} = \pi'_{ij}\prod_{k=1}^{j-1}(1-\pi'_{ik})$ where, $\pi'_{ij}{\sim}\text{ Beta }(\alpha_0\gamma_j, \alpha_0(1-\sum_{k=1}^{j}\gamma_k))$. Hence the model can be written as
\begin{align}
    \undertilde{N_i}|\undertilde{\pi_i} & \overset{\text{ind}}{\sim} F(\undertilde{\pi_i}) \qquad i=1,2,\ldots
\end{align}
where the prior distribution on the $\undertilde{\pi_i}$ is given by 
\begin{align}
        \undertilde{\pi_i} & = (\pi_{ij})_{j=1}^{\infty} \sim \text{DP}(\alpha_0, \undertilde{\gamma})\\
        \implies \pi_{ij} & = \pi'_{ij}\prod_{k=1}^{j-1}(1-\pi'_{ik})
\end{align}
where,
\begin{align}
   \pi'_{ij}|(\gamma_1,\gamma_2,\dots,\gamma_j){\sim}\text{ Beta }(\alpha_0\gamma_j, \alpha_0(1-\sum_{k=1}^{j}\gamma_k)) \qquad i,j=1,2,\ldots
\end{align} 
where the distribution of the hyper-parameter $\undertilde{\gamma}$ is as follows
\begin{align*}
    \gamma_j = \nu_{j} \prod_{k=1}^{j-1}(1-\nu_{k}) \qquad j=1,2,\ldots \text{ and }
    \sum_{j=1}^{\infty}\gamma_{j}=1
\end{align*}
where,
\begin{align*}
    \nu_{j} \overset{i.i.d}{\sim} \text{Beta}(\alpha,\beta) \qquad j=1,2,\ldots
\end{align*}
\subsection{Choice of Posterior Sampling Algorithm}
For Dirichlet Process Mixture Models, \cite{ishwaran2001gibbs} showed that the blocked Gibbs sampling technique to obtain samples from the posterior distribution performed better in terms of both mixing and scalability than Chinese Restaurant Process based collapsed samplers. In this algorithm the infinite dimensional Dirichlet process prior was approximated by its finite version which in turn made the number of model parameters finite (\cite{ishwaran2002exact}). The parameters are then updated in blocks from simple multivariate distributions. \cite{das2024blocked} have developed a blocked Gibbs sampler that samples from the posterior distribution of the truncated or finite version of the Hierarchical Dirichlet Process, which produces statistically stable results and is scalable with good mixing. We extend this idea to our problem by using blocked Gibbs sampler to sample from the truncated approximation of the Generalized Hierarchical Stick Breaking Process prior. 
\section{Finite Version of the Model}
In order to obtain the posterior distribution of the parameters, we work with the finite version of the model. Here we truncate the TPM by taking the first $d$ states only. In the finite case, the distribution of the number of observations in the $i$th row of the Transition Probability Matrix, $\undertilde{N_i}=(N_{i1},N_{i2},\ldots,N_{id})$ given $\pi_{i2},\ldots,\pi_{id}$ will follow a multinomial distribution. That is, defining $n_{i}=\sum_{j=1}^d N_{ij}$, the model can be written as
\begin{align*}
    \gamma_j & = \nu_{j} \prod_{k=1}^{j-1}(1-\nu_{k}) \qquad j=1,2,\ldots,d-1 \text{ and } \gamma_d = \prod_{k=1}^{d-1}(1-\nu_{k}) \text{ such that }
    \sum_{j=1}^{d}\gamma_{j}=1\\
    \nu_{j} & \overset{i.i.d}{\sim} \text{Beta}(\alpha,\beta)\\
    \pi_{ij} & = \pi'_{ij}\prod_{k=1}^{j-1}(1-\pi'_{ik}) \qquad j=1,2,\ldots,d-1 \text{ and } \pi_{id} = \prod_{k=1}^{d-1}(1-\pi'_{ik}) \text{ such that }
    \sum_{j=1}^{d}\pi_{ij}=1\\
    \pi'_{ij} & \sim \text{ Beta }(\alpha_0\gamma_j, \alpha_0(1-\sum_{k=1}^{j}\gamma_k)) \qquad \forall i=1,2,\ldots,d\\
    & N_{i1},N_{i2},\ldots,N_{id}|\pi_{i1}, \pi_{i2},\ldots,\pi_{id} \sim \text{Multinomial}(n_{i};\pi_{i1},\pi_{i2},\ldots,\pi_{id})
\end{align*}
\subsection{Distribution of the Model Parameters in the Finite Case}
\begin{lemma}\label{lemma2}
    In the finite case, $\undertilde{\gamma}_{d-1} = (\gamma_1,\gamma_2,\ldots,\gamma_{d-1})$ follows a Generalized Dirichlet Distribution, that is,
    \[
        \undertilde{\gamma}_{d-1} \sim \text{GD}_{d-1}(\underbrace{\alpha,\alpha,\ldots,\alpha}_\textrm{d-1 \text{ times}};\underbrace{\beta,\beta,\ldots,\beta}_\textrm{d-1 \text{ times}})
    \]
    Here the p.d.f of $\undertilde{\gamma}$ is specified by
    \[
        \frac{1}{[B(\alpha,\beta)]^{d-1}}(1-\gamma_1-\ldots-\gamma_{d-1})^{\beta-1}\prod_{i=1}^{d-1}{\gamma_i}^{\alpha-1}(1-\gamma_1-\ldots-\gamma_{i-1})^{-\alpha}
    \]
    where, $\sum_{i=1}^{d-1}\gamma_i < 1$. Moreover,
    given $\undertilde{\gamma}$, each $\undertilde{\pi_i}$ is independently distributed as Dirichlet($\alpha_0\undertilde{\gamma}$), that is,
    \[
        \undertilde{\pi_i}|\undertilde{\gamma} \overset{i.i.d}{\sim} \text{Dirichlet}(\alpha_0\undertilde{\gamma}) \qquad{\forall i=1,2,\ldots,d}
    \]
\end{lemma}
Proof is provided in Appendix \ref{Appendix-B.4}.

\subsection{Sampling from the Posterior Distribution}
In order to construct a blocked Gibbs sampler for the Generalized Stick Breaking Process we will be working with the finite approximation of the model and take the prior distribution of the shared concentration parameter $\alpha_0$ to be a gamma prior. 
Suppose that we have observed states $1$ to $d$ in our data. 
Let the prior distribution of $\alpha_0$ be Gamma$(a_0,b_0)$. This choice of prior of $\alpha_0$ is taken deliberately to make simplification of the posterior distribution easier in the following calculations.
Then the joint posterior distribution of the parameters $(\bm{\pi},\undertilde{\gamma},\alpha_0,\alpha,\beta)$ where $\alpha_0 \sim$ Gamma$(a_0,b_0)$ is given by
\begin{align*}
    P(\bm{\pi},\undertilde{\gamma},\alpha_0, \alpha, \beta|\bm{n})
    & \propto P(\bm{n}|\bm{\pi})P(\bm{\pi}|\alpha_0,\undertilde{\gamma})P(\undertilde{\gamma}|\alpha,\beta)P(\alpha_0) \\
    & \propto \left(\prod_{i=1}^d \pi_{i1}^{n_{i1}}\pi_{i2}^{n_{i2}}\ldots\pi_{i,d-1}^{n_{i,d-1}}\pi_{id}^{n_{id}} \right) \frac{1}{[B(\alpha,\beta)]^{d-1}} \gamma_d^{\beta-1} \left(\prod_{j=1}^{d-1} \gamma_j^{\alpha-1}(1-\gamma_1-\ldots-\gamma_{j-1})^{-\alpha} \right) \mathbb{1}_{[\gamma \in \mathcal{S}^{d-1}]}\\
    & \quad \prod_{i=1}^d \frac{\Gamma(\alpha_0)}{\prod_{j=1}^d\Gamma(\alpha_0\gamma_j)} \left(\prod_{j=1}^d \pi_{ij}^{\alpha_0 \gamma_j-1} \right)\mathbb{1}_{[\pi_i \in \mathcal{S}^{d-1}]} \alpha_0^{a_0-1}e^{-b_0 \alpha_0}  \mathbb{1}_{[\alpha_0>0]}\\
\end{align*}
where $\mathcal{S}^{d-1}$ denotes the $d$-dimensional simplex. The full conditional distributions for the parameters are worked out from their joint posterior distribution, and the blocked parameter updates are given below. 
\begin{enumerate}
    \item Sampling $\bm{\pi} $ 
    \[
        \pi_i|\undertilde{\gamma},\alpha_0 \sim \text{Dirichlet}(\undertilde{n_i} + \alpha_0\undertilde{\gamma}), \qquad i=1,2,\ldots,d
    \]
where, $n_i = (n_{i1},n_{i2},\ldots,n_{id})$
We want to sample the posterior distribution of $\undertilde{\gamma}$
    \item Sampling $\undertilde{\gamma}$
    \begin{align*}
        P(\undertilde{\gamma}|\bm{\pi},\alpha_0) \propto \prod_{j=1}^d \frac{1}{\Gamma(\alpha_0 \gamma_j)^d} \prod_{i=1}^d \pi_{ij}^{\alpha_0\gamma_j - 1} \gamma_d^{\beta-1} \prod_{j=1}^{d-1} \gamma_j^{\alpha-1}(1-\gamma_1-\ldots-\gamma_{j-1})^{-\alpha}
    \end{align*}
    \item Sampling $\alpha_0$
    \begin{align*}
        P(\alpha_0|\undertilde{\gamma},\bm{\pi}) \propto \prod_{i=1}^d \frac{\Gamma(\alpha_0)}{\prod_{j=1}^d\Gamma(\alpha_0\gamma_j)} \left(\prod_{j=1}^d \pi_{ij}^{\alpha_0 \gamma_j-1} \right)\mathbb{1}_{[\pi_i \in \mathcal{S}^{d-1}]} \alpha_0^{a_0-1}e^{-b_0 \alpha_0}  \mathbb{1}_{[\alpha_0>0]}
    \end{align*}
\end{enumerate}
From the full conditional distributions it is observed that for each $i=1,2,\ldots,d$, the random vector $\pi_i$ has a closed form conjugate update but the posteriors of the parameters $\undertilde{\gamma}$ and $\alpha_0$ do not have any standard form. Since $\alpha_0$ is a one-dimensional variable, we can easily draw samples from its posterior distribution using the Metropolis-Hastings algorithm. However, the global parameter $\undertilde{\gamma}$, lies on a $d-1$-dimensional simplex and the value of $d$ may be very large in practical scenarios. In such situations, the acceptance rate in Metropolis Hastings algorithm becomes negligible, leading to high rejection rate of most of the proposed samples which will lead to poor convergence and low mixing. In order to overcome this problem, we make suitable transformations and introduce additional latent or auxiliary random variables in such a way that the augmented posterior distribution can be expressed as a product of independent random variables making it easier and more efficient to draw samples from it.
\subsection{Sampling of the Global Parameter}
Since our objective is to estimate the infinite dimensional transition probability matrix and we are working with a finite approximation of the said matrix, hence the dimension $d$ of the $\undertilde{\gamma}$ vector may be quite large in practice which will lead to difficulty in sampling from its posterior distribution. To overcome this problem, we to express $\undertilde{\gamma}$ as a normalization of non-negative random variables such that
\[
\gamma_j = \frac{t_j}{\sum_{k=1}^d t_k}, \qquad t_j>0 \quad \forall j=1(1)d 
\]
and, then introduce auxiliary variables in such a way that the posterior density function becomes a product of independent random variables which makes it easier to use some MCMC algorithm to sample from it.
Similar technique has been used by \cite{das2024blocked} to simplify the posterior distribution of the global weights of a hierarchical Dirichlet process. \\
The introduction of the random vector $\undertilde{t}$ causes the prior distribution of $\undertilde{\pi_i}$ to be expressed as
\[
    \undertilde{\pi_i}|\undertilde{t}, \alpha_0 \sim \text{ Dir}(\undertilde{t})
\]
Now the joint prior distribution of $(\bm{\pi},\undertilde{\gamma}, \alpha_0)$ is given by 
\begin{align*}
    P(\bm{\pi},\undertilde{\gamma}, \alpha_0)
    & = \gamma_d^{\beta-1} \prod_{j=1}^{d-1} \gamma_j^{\alpha-1}(1-\gamma_1-\ldots-\gamma_{j-1})^{-\alpha} \mathbb{1}_{[\undertilde{\gamma} \in \mathcal{S}^{d-1}]}\\
    & \prod_{i=1}^d \frac{\Gamma(\alpha_0)}{\prod_{j=1}^d\Gamma(\alpha_0\gamma_j)} \pi_{ij}^{\alpha_0 \gamma_j-1} \mathbb{1}_{[\undertilde{\pi_i} \in \mathcal{S}^{d-1}]} \alpha_0^{a_0-1}e^{-b_0 \alpha_0}  \mathbb{1}_{[\alpha_0>0]}
\end{align*}
Integrating over $\alpha_0$, the joint distribution of $(\bm{\pi},\undertilde{\gamma})$ becomes 
\begin{align*}
    P(\bm{\pi},\undertilde{\gamma})
    & = \int_0^\infty \alpha_0^{d-\beta+a_0-1} \Gamma(\alpha_0)^d (\alpha_0\gamma_d)^{\beta-1} \prod_{j=1}^d \frac{e^{-b_0\alpha_0\gamma_j}}{\Gamma(\alpha_0 \gamma_j)^d} (\alpha_0\gamma_j)^{\alpha-1}(\alpha_0-\alpha_0\gamma_1-\ldots-\alpha_0\gamma_{j-1})^{-\alpha} \\
    & \prod_{i=1}^d \pi_{ij}^{\alpha_0\gamma_j-1}\mathbb{1}_{[\undertilde{\pi_i} \in \mathcal{S}^{d-1}]} 
    \mathbb{1}_{[\undertilde{\gamma} \in \mathcal{S}^{d-1}]} d\alpha_0
\end{align*}
It is seen that $\sum_{k=1}^d t_k = \alpha_0$. Expressing this quantity as $t$, that is $t = \sum_{k=1}^d t_k = \alpha_0$, the joint prior density of $(\bm{\pi},\undertilde{\gamma})$ becomes
\begin{align*}
    P(\bm{\pi},\undertilde{\gamma})
    & = \int_0^\infty t^{d-\beta+a_0-1} \Gamma(t)^d (t\gamma_d)^{\beta-1} \prod_{j=1}^d \frac{e^{-b_0t\gamma_j}}{\Gamma(t \gamma_j)^d} (t\gamma_j)^{\alpha-1}(t-t\gamma_1-\ldots-t\gamma_{j-1})^{-\alpha} \\
    & \prod_{i=1}^d \pi_{ij}^{t\gamma_j-1}\mathbb{1}_{[\undertilde{\pi_i} \in \mathcal{S}^{d-1}]} 
    \mathbb{1}_{[\undertilde{\gamma} \in \mathcal{S}^{d-1}]} dt
\end{align*}
For simplification we take $a_0$ to be $\beta$, that is the distribution $\alpha_0$ is assumed to follow Gamma$(\beta,b_0)$. Then using lemma 1 of \cite{das2024blocked}, we get,
\begin{align*}
    P(\bm{\pi},\undertilde{t}) & = \Gamma(t)^d (t_d)^{\beta-1} \prod_{j=1}^d \frac{e^{-b_0t_j}}{\Gamma(t_j)^d} (t_j)^{\alpha-1}(t-t_1-\ldots-t_{j-1})^{-\alpha} \\
    & \prod_{i=1}^d \pi_{ij}^{t_j-1}\mathbb{1}_{[\undertilde{\pi_i} \in \mathcal{S}^{d-1}]} 
    \mathbb{1}_{[\undertilde{t} > \undertilde{0}]}\\
    P(\bm{\pi},\undertilde{t}) & = \Gamma(t)^d (t_d)^{\beta-1} \prod_{j=1}^d \frac{e^{-b_0t_j}}{\Gamma(t_j)^d} (t_j)^{\alpha-1}(t_j+t_{j+1}+\ldots+t_{d})^{-\alpha} \prod_{i=1}^d \pi_{ij}^{t_j-1}\mathbb{1}_{[\undertilde{\pi_i} \in \mathcal{S}^{d-1}]} 
    \mathbb{1}_{[\undertilde{t} > \undertilde{0}]}
\end{align*}
Here the terms $\Gamma(t)^d$ and $(t_j+t_{j+1}+\ldots+t_{d})^{-\alpha}$ prevent the posterior density of $\undertilde{t}$ to be written as a product of independent densities. In order to eliminate this dependency we incorporate suitably chosen auxiliary random variables $u_1,u_2,\ldots,u_d,$ and $w_1,w_2,\ldots,w_d$  such that
\begin{align*}
    \Gamma(t)^d 
    = \Gamma(\sum_{j=1}^d t_j)^d
    & = \prod_{i=1}^d \left\{ \int_0^\infty e^{-u_i} u_i^{\sum_{j=1}^d t_j - 1} du_i \right\}\\
    \prod_{j=1}^d (t_j+t_{j+1}+\ldots+t_{d})^{-\alpha} 
    & =  \Gamma(\alpha)^d  \prod_{j=1}^d  \left\{ \int_0^\infty e^{-(t_j+t_{j+1}+\ldots+t_{d})w_j} w_j^{\alpha-1} dw_j \right\}
\end{align*}
Then the augmented posterior of $(\bm{\pi},\undertilde{t},\undertilde{u},\undertilde{w})$ becomes
    \begin{align*}
    P(\bm{\pi},\undertilde{t},\undertilde{u},\undertilde{w})
    \propto & \prod_{i=1}^d \left\{e^{-u_i} u_i^{\sum_{j=1}^d t_j - 1} \mathbb{1}_{[u_i > 0]} \right\}\\
    & \Gamma(\alpha)^d (t_d)^{\beta-1} \prod_{j=1}^d \frac{e^{-b_0 t_j}}{\Gamma(t_j)^d} (t_j)^{\alpha-1}e^{-(t_j+t_{j+1}+\ldots+t_{d})w_j} w_j^{\alpha-1} \mathbb{1}_{[\undertilde{w} > \undertilde{0}]}\\
    & \prod_{i=1}^d \pi_{ij}^{t_j-1}\mathbb{1}_{[\undertilde{\pi_i} \in \mathcal{S}^{d-1}]} 
    \mathbb{1}_{[\undertilde{t} > \undertilde{0}]}
\end{align*}
The full conditional posteriors of $\undertilde{u}, \undertilde{t}$ and $\undertilde{w}$ are given by
\begin{align}
    P(\undertilde{u}|t) 
    & \propto \prod_{i=1}^d e^{-u_i} u_i^{\sum_{j=1}^d t_j - 1} \mathbb{1}_{[u_i > 0]}\\
    P(\undertilde{w}|\alpha, \undertilde{t})
    & \propto \prod_{j=1}^d e^{-(t_j+t_{j+1}+\ldots+t_{d})w_j} w_j^{\alpha-1}\mathbb{1}_{[w_j > 0]}\\
    P(\undertilde{t}|\bm{\pi},\undertilde{u},\undertilde{w})
     & \propto \prod_{j=1}^d f_j(t_j)
\end{align}
where,
\begin{align*}
    f_j(t) \propto \frac{1}{\Gamma(t)^d} e^{-(b_0 + \sum_{k=1}^j w_k - \sum_{i=1}^d \log \pi_{ij} - \sum_{i=1}^d \log u_i)t} t^{\delta_j-1}\mathbb{1}_{[t > 0]}, \qquad \delta_j = \begin{cases}
        \alpha\text{ if } j=1,2,\cdots,d-1\\
        \beta \text{ if } j=d
    \end{cases}
\end{align*}
Thus $\undertilde{u}|t$ is a vector of independent Gamma($t$,1) random variables and $w_j|\alpha, \undertilde{t}$ follows Gamma($\alpha,\sum_{k=j}^d t_k)$ for $j=1,\dots,d$ independently for each $j$. $f_j$ is a tilted Gamma density with parameters $d, \delta_j, B_j = b_0 + \sum_{k=1}^j w_k - \sum_{i=1}^d \log \pi_{ij} - \sum_{i=1}^d \log u_i$. The final parameter updates are provided in Appendix \ref{Appendix-A}.\\
The samples from this tilted gamma density can be obtained using a rejection sampler. The construction of a suitable target density for this rejection sampler is given by \cite{das2024blocked}. The rejection sampling algorithm is provided in Appendix \ref{Appendix-B}. 

\section{Simulation Results}
To test the validity of our model, we generate a Markov Chain $(x_1,\dots,x_n)$ using the following steps.
\begin{itemize}
    \item Start from State 0 (We may also use a random start of any natural number)
    \item The probability of going to the $j^{th}$ step when we are at the $i^{th}$ step is calculated as $q_{i}^jp_i$, that is the $i^{th}$ row follows a Geometric($p_i$) distribution.
\end{itemize}
We consider the following two functional forms of $p_i$, 
\begin{enumerate}
    \item $p_i = \frac{1}{\log(x_i)+10}$, and
    \item  $pi = \frac{1}{\log(\log(x_i)+100)}$
\end{enumerate}
A sample Markov Chain of length 200,000 is generated in each case. The estimated transition probability matrix is then compared with the true transition probability matrix where the estimation is performed using three methods, the maximum likelihood estimator, the Hierarchical Stick-Breaking prior and the Generalized Hieararchical Stick-Breaking Prior taking different values of the hyper-parameters $\alpha, \beta$ and $b_0$. The Hierarchical Stick-Breaking Prior is a special case of the Generalized Hierarchical Stick-Breaking Prior with $\alpha = 1$.

We take $M=2000$ posterior samples after a burn-in of $1000$ samples. To construct the cover density of the tilted Gamma distribution, we use $2N+2$ knot points. Here we have taken $N=2$ for running the simulations. The mean absolute errors (MAE) are computed truncating the true transition probability matrix till the maximum number of observed states in the sample. 

The results of the simulation are presented in Table 1 and Table 2. Table 1 corresponds to the first function of $p_i$, that is $p_i = \frac{1}{\log(x_i)+10}$ and Table 2 corresponds to that of the second function, that is $p_i = \frac{1}{\log(\log(x_i)+100))}$. In both cases, it is observed that the Generalized Hierarchical Stick-Breaking Prior consistently achieves the lowest Mean Absolute Error (MAE) for several combinations of the hyper-parameters. 
\begin{table}[ht!]
\centering
\caption{Mean Absolute Errors ($\times 100$) under the functional form $p_i = \frac{1}{\log(x_i) + 10}$ corresponding to the Maximum Likelihood Estimator (MLE), the Hierarchical Stick-Breaking Prior (HSBP), and the Generalized Hierarchical Stick-Breaking Prior (GHSBP).}
\begin{tabular}{l|p{1.2cm} p{1.2cm} c|c}
\toprule
\textbf{Method} & \textbf{$\alpha$} & \textbf{$\beta$} & \textbf{$b_0$} & \textbf{Mean Absolute Error $\times$ 100} \\
\midrule
\textbf{MLE} & -- & -- & -- & 1.727 \\
\midrule
\multirow{6}{*}{\textbf{Hierarchical Stick-Breaking Prior}}
    & 1 & 0.5 & 10 & 1.077 \\
    & 1 & 1 & 10 & 1.080 \\
    & 1 & 2 & 10 & 1.079 \\
    & 1 & 2 & 25 & 1.184 \\
    & 1 & 2 & 50 & 1.260 \\
    & 1 & 5 & 50 & 1.261 \\
\midrule
\multirow{12}{*}{\textbf{Generalized Hierarchical Stick-Breaking Prior}}
    & 2 & 0.5 & 10 & 1.076 \\
    & 2 & 2 & 10 & 1.070 \\
    & 3 & 2 & 10 & 1.069 \\
    & 3 & 1 & 10 & 1.068 \\
    & 5 & 2 & 10 & 1.074 \\
    & 2 & 2 & 25 & 1.144 \\
    & 5 & 2 & 25 & 1.110 \\
    & 2 & 2 & 50 & 1.219 \\
    & 3 & 2 & 50 & 1.191 \\
    & 2 & 5 & 50 & 1.215 \\
    & 3 & 5 & 50 & 1.190 \\
    & 5 & 5 & 50 & 1.161 \\
\bottomrule
\end{tabular}
\end{table}
\begin{table}[ht!]
\centering
\caption{Mean Absolute Errors ($\times 100$) under the functional form $p_i = \frac{1}{\log(\log(x_i) + 100)}$ corresponding to the Maximum Likelihood Estimator (MLE), the Hierarchical Stick-Breaking Prior (HSBP), and the Generalized Hierarchical Stick-Breaking Prior (GHSBP).}
\begin{tabular}{l|p{1.2cm} p{1.2cm} c|c}
\toprule
\textbf{Method} & \textbf{$\alpha$} & \textbf{$\beta$} & \textbf{$b_0$} & \textbf{Mean Absolute Error $\times$ 100} \\
\midrule
\textbf{MLE} & -- & -- & -- & 1.829 \\
\midrule
\multirow{5}{*}{\textbf{Hierarchical Stick-Breaking Prior}} 
    & 1 & 1 & 10 & 2.284\\
    & 1 & 0.5 & 10 & 2.274\\
    & 1 & 0.1 & 5 & 2.199 \\
    & 1 & 0.1 & 2 & 2.155 \\
    & 1 & 0.001 & 0.1 & 2.115 \\
\midrule
\multirow{13}{*}{\textbf{Generalized Hierarchical Stick-Breaking Prior}}
    & 20 & 1 & 10 &  1.784\\
    & 30 & 1 & 10 & 1.756 \\
    & 50 & 1 & 10 & 1.726 \\
    & 10 & 0.5 & 10 & 1.864 \\
    & 15 & 0.5 & 10 & 1.809 \\
    & 30 & 0.5 & 10 & 1.751 \\
    & 50 & 0.5 & 10 & 1.756 \\
    & 10 & 0.1 & 5 & 1.866 \\
    & 20 & 0.1 & 5 & 1.767 \\
    & 20 & 0.1 & 2 & 1.788 \\
    & 10 & 0.001 & 2 &  1.830\\
    & 20 & 0.001 & 2 &  1.783\\
\bottomrule
\end{tabular}
\end{table}
\clearpage
	\bibliographystyle{plainnat}
	\setcitestyle{numbers}
	\bibliography{mybib}
\newpage
\begin{appendices}
    \appendix
\section{Final Parameter Updates} \label{Appendix-A}
\begin{enumerate}
    \item Sampling $\bm{\pi} $ 
    \[
        \pi_i|\undertilde{t} \sim \text{Dirichlet}(\undertilde{n_i} + \undertilde{t}), \qquad i=1,2,\ldots,d
    \]
    where, $n_i = (n_{i1},n_{i2},\ldots,n_{id})$
    \item Sampling $\undertilde{u}$
    \begin{align*}
        P(\undertilde{u}|t) & \propto \prod_{i=1}^d e^{-u_i} u_i^{\sum_{j=1}^d t_j - 1} \mathbb{1}_{[u_i > 0]}
    \end{align*}
    That is, 
    \[
    u_j|t \overset{ind}{\sim}  \text{ Gamma}(t,1) \text{ for all } j=1,\dots,d
    \]
    \item Sampling $\undertilde{w}$
    \begin{align*}
         P(\undertilde{w}|\alpha, \undertilde{t})
        & \propto \prod_{j=1}^d e^{-(t_j+t_{j+1}+\ldots+t_{d})w_j} w_j^{\alpha-1}\mathbb{1}_{[w_j > 0]}\\
    \end{align*}
    That is, 
    \[
    w_j|\alpha, \undertilde{t} \overset{ind}{\sim} \text{ Gamma}(\alpha,\sum_{k=j}^d t_k) \text{ for all } j=1,\dots,d
    \]
    \item Sampling $\undertilde{t}$
    \begin{align*}
    P(\undertilde{t}|\bm{\pi},\undertilde{u},\undertilde{w}) & \propto \prod_{j=1}^d f_j(t_j)
    \end{align*}
    where,
    \begin{align*}
        f_j(t) \propto \frac{1}{\Gamma(t)^d} e^{-(b_0 + \sum_{k=1}^j w_k - \sum_{i=1}^d \log \pi_{ij} - \sum_{i=1}^d \log u_i)t} t^{\delta_j-1}\mathbb{1}_{[t > 0]}, \qquad \delta_j = \begin{cases}
        \alpha\text{ if } j=1,2,\cdots,d-1\\
        \beta \text{ if } j=d
        \end{cases}
    \end{align*}
    That is, $f_j$ is a tilted Gamma density with parameters $d, \delta_j, B_j = b_0 + \sum_{k=1}^j w_k - \sum_{i=1}^d \log \pi_{ij} - \sum_{i=1}^d \log u_i$ and $\alpha_0 = t$.
\end{enumerate}

\section{Proof of Auxiliary Results}
\label{Appendix-B}
\subsection{Proof of Lemma~\ref{lemma0}}\label{Appendix-B.1}
\begin{proof}[Proof:]
    Expressing $x_j$'s in terms of $y_j$'s, we get,
    \begin{align*}
        x_1 & = y_1\\
        x_2 & = \frac{y_2}{1-y_1}\\
        x_3 & = \frac{y_3}{1-y_1-y_2}\\
        & \vdotswithin{=}\\
        x_{d} & = \frac{y_{d}}{1-y_1-y_2-\ldots-y_{d-1}}
    \end{align*}
    Given each $x_j$ is independently and identically distributed as Beta$(\alpha_j,\beta_j)$, the joint distribution of $x_j,j=1,2,\ldots,d-1$ is given by 
        \begin{align*}
            f(x_1,\ldots,x_{d-1}) & = \prod_{j=1}^{d-1} \frac{1}{B(\alpha_j,\beta_j)} x_j^{\alpha-1} (1-x_j)^{\beta-1}, \qquad 0<x_j<1 \forall j=1,\dots,d-1, \sum_{j=1}^{d-1} x_j < 1
        \intertext{The Jacobian of transformation is calculated as}
            |\textbf{J}| & = \left|\pdv{(x_1,\ldots,x_{d-1})}{(y_1,\ldots,y_{d-1})}\right| = \prod_{j=1}^{d-2} \frac{1}{1-y_1-\ldots-y_j}
        \intertext{Hence the probability density function of $(y_1,y_2,\ldots,y_{d-1})$ will be} 
            f(y_1,y_2,\ldots,y_{d-2}) & = \prod_{j=1}^{d-1} \frac{1}{B(\alpha_j,\beta_j)} \frac{y_j^{\alpha_j - 1}(1-y_1-\ldots-y_j)^{\beta_j-1}}{(1-y_1-\ldots-y_{j-1})^{\alpha_j-\beta_j-2}} \prod_{j=1}^{d-2}\frac{1}{1-y_1-\ldots-y_j}
        \intertext{Or,}
             f(y_1,y_2,\ldots,y_{d-1}) & = y_{d-1}^{\alpha_d-1} (1-y_1-y_2-\ldots-y_{d-1})^{\beta_{d-1}} \prod_{j=1}^{d-2} \frac{1}{B(\alpha_j,\beta_j)} y_j^{\alpha_i-1}(1-y_1-y_2-\ldots-y_j)^{\beta_j - \alpha_{j+1} - \beta{j+1}}
        \intertext{where,}
             & \qquad 0<y_j<1 \forall j=1,\ldots,d-1 \text{ and } \sum_{j=1}^{d-1} y_j < 1
        \end{align*}
    \end{proof}
\subsection{Proof of Lemma~\ref{lemma1}} \label{Appendix-B.2}
\begin{proof}[Proof:]
      If a random vector $\undertilde{Y}=(Y_1,Y_2,\ldots,Y_k)$ is such that
    \[
        Y_i \overset{\text{ind}}{\sim} \text{Gamma}(\alpha_i) \qquad i=1,2,\ldots,k
    \]
    then the random vector $\undertilde{X}=(X_1,X_2,\ldots,X_k)=\undertilde{Y}=\frac{1}{\sum_{i=1}^k Y_i}(Y_1,Y_2,\ldots,Y_k)$ follows a Dirichlet distribution with parameters $(\alpha_1,\alpha_2,\ldots,\alpha_k)$, that is,
    \[
        \undertilde{X}=(X_1,X_2,\ldots,X_k) \sim \text{Dirichlet}(\alpha_1,\alpha_2,\ldots,\alpha_k)
    \]
    Similarly,
    \[
       \frac{1}{\sum_{i=2}^k Y_i}(Y_2,Y_3,\ldots,Y_k) \sim \text{Dirichlet}(\alpha_2,\alpha_3,\ldots,\alpha_k)
    \]
    Now,
    \begin{align}
        \frac{1}{\sum_{i=2}^k Y_i}(Y_2,Y_3,\ldots,Y_k) 
        & = \frac{1}{\sum_{i=1}^k Y_i-Y_1}(Y_2,Y_3,\ldots,Y_k)\\
        & = \frac{1}{1-\frac{Y_1}{\sum_{i=1}^k Y_i}}\left(\frac{Y_2}{\sum_{i=1}^k Y_i},\frac{Y_3}{\sum_{i=1}^k Y_i},\ldots,\frac{Y_k}{\sum_{i=1}^k Y_i}\right)\\
        & = \frac{1}{1-X_1}(X_2,X_3,\ldots,X_k)
    \end{align}
    which implies,
    \begin{align}
        \frac{1}{1-X_1}(X_2,X_3,\ldots,X_k)  \sim \text{Dirichlet}(\alpha_2,\alpha_3,\ldots,\alpha_k)
    \end{align}
\end{proof}
\subsection{Proof of Proposition~\ref{GSBP_proof}}\label{Appendix-B.3}
\begin{proof}[]
     Let $(A_1=I_1,A_2=I_2,\ldots,A_r=I_r)$ be a finite partition $\Theta = \mathbb{N}$, the set of positive integers. Then, using $\ref{eq0}$, the Dirichlet process defined in Equation $\ref{eq6}$ implies that for each $j$,
     \begin{align}
        \left(G_j(A_1),G_j(A_2),\ldots,G_j(A_r)\right) & \sim \text{Dirichlet}\left(\alpha_0 G_0(A_1),\alpha_0 G_0(A_2),\ldots,\alpha_0 G_0(A_r)\right) \nonumber \\ 
        \implies \left(\sum_{j \in I_1} \pi_{ij},\ldots,\sum_{j \in I_r} \pi_{ij}\right) & \sim \text{Dirichlet} \left(\alpha_0\sum_{j \in I_1}\beta_j,\ldots,\alpha_0\sum_{j \in I_r}\beta_j\right)
     \end{align}
    To derive (\ref{eq1}), we make a partition $(\{1,2,\dots,j-1\},\{j\},\{j+1,j+2\,\dots\})$ of $\mathbb{N}$ such that,
    \[
        \left(\sum_{k=1}^{j-1} \pi_{ik},\pi_{ij},\sum_{k=j+1}^{\infty} \pi_{ik}\right) \sim \text{Dirichlet}\left(\alpha_0 \sum_{k=1}^{j-1}\gamma_k,\alpha_0\gamma_j,\alpha_0\sum_{k=j+1}^{\infty}\gamma_k\right)
    \]
    Then by removing the first element and using Lemma \ref{lemma1},
    \begin{align}
        \frac{1}{1-\sum_{k=1}^{j-1}\pi_{ik}}\left(\pi_{ij},\sum_{k=j+1}^{\infty} \pi_{ik}\right) \sim \text{Dirichlet}\left(\alpha_0\gamma_j,\alpha_0\sum_{k=j+1}^{\infty}\gamma_k \right) \label{eq3}
    \end{align}
    Equation $\ref{eq2}$ implies that $\pi'_{ij}=\frac{\pi_{ij}}{1-\sum_{k=1}^{j-1}\pi_{ik}}$. Again, $\frac{\sum_{k=j+1}^{\infty} \pi_{ik}}{1-\sum_{k=1}^{j-1}\pi_{ik}}=\frac{1-\sum_{k=1}^{j} \pi_{ik}}{1-\sum_{k=1}^{j-1}\pi_{ik}}=1-\pi'_{ij}$, and $\sum_{k=j+1}^{\infty}\gamma_k = 1-\sum_{k=1}^j \gamma_k$. Hence, from (\ref{eq3}),
    \begin{align*}
       (\pi'_{ij},1-\pi'_{ij}) & \sim \text{Dirichlet}\left(\alpha_0\gamma_j,\alpha_0\sum_{k=1}^j(1-\gamma_k) \right).\\
       \implies   \pi'_{ij} & \sim \text{ Beta }(\alpha_0\gamma_j, \alpha_0(1-\sum_{k=1}^{j}\gamma_k)) \qquad i,j=1,2,\ldots
    \end{align*}
\end{proof}
\subsection{Proof of Lemma~\ref{lemma2}}\label{Appendix-B.4}
\begin{proof}[Proof:]
    Given that each $\nu_j$ is independently and identically distributed as a Beta$(\alpha,\beta)$ and the random vector $\undertilde{\gamma} =(\gamma_1,\gamma_2,\ldots,\gamma_d)$ is defined as $\gamma_j = \nu_{j} \prod_{k=1}^{j-1}(1-\nu_{k})$ for all $j=1,2,\ldots,d-1$ and $\gamma_d = \prod_{k=1}^{d-1}(1-\nu_{k})$ such that $\sum_{j=1}^{d}\gamma_{j}=1$, then it follows from the construction of Generalized Dirichlet Distribution (Lemma \ref{lemma0}) that 
    \begin{align*}
        \undertilde{\gamma}_{d-1} = (\gamma_1,\gamma_2,\ldots,\gamma_{d-1}) \sim \text{GD}_{d-1}(\underbrace{\alpha,\alpha,\ldots,\alpha}_\textrm{d-1 \text{ times}};\underbrace{\beta,\beta,\ldots,\beta}_\textrm{d-1 \text{ times}})
    \end{align*}
    Similarly, given that for each $i$, the random variables $\pi'_{ij}$'s are independently distributed as Beta$(\alpha_0\gamma_j, \alpha_0(1-\sum_{k=1}^{j}\gamma_k))$ with $\pi_{ij} = \pi'_{ij}\prod_{k=1}^{j-1}(1-\pi'_{ik})$ for all $j=1,2,\ldots,d-1$ and $\pi_{id} = \prod_{k=1}^{d-1}(1-\pi'_{ik})$ such that $\sum_{j=1}^{d}\pi_{ij}=1$, then again by $\ref{lemma0}$, for all $i=1,2,\ldots,d$,
    \begin{align*}
        \pi_{i1},\ldots,\pi_{id-1} & \sim \text{GD} \left(\alpha_0\gamma_1,\ldots,\alpha_0\gamma_{d-1};\alpha_0(1-\gamma_1),\ldots,\alpha_0(1-\sum_{k=1}^{d-1}\gamma_k)\right)
    \end{align*}
    Hence the probability density function of $(\pi_{i1},\ldots,\pi_{id-1})$ becomes
    \begin{align*}
        f(\pi_{i1},\ldots,\pi_{id-1}) & \propto \pi_{id-1}^{\alpha_0\gamma_{d-1}}(1-\pi_{i1}-\ldots-\pi_{id-1})^{\alpha_0(1- \sum_{k=1}^{d-1}\gamma_k)}\prod_{j=1}^{d-2} \pi_{ij}^{\alpha_0 \gamma_j}(1-\pi_{i1}-\ldots-\pi_{ij})^{\alpha_0(1-\sum_{k=1}^{j}\gamma_k) - \alpha_0\gamma_{j+1}-\sum_{k=1}^{j+1}\gamma_k}
    \end{align*}
    Now, the denominator corresponding to the term $1-\pi_{i1}-\ldots-\pi_{ij}$ becomes 0 for all $j=1,\ldots,d-2$. Moreover, given that $1-\pi_{i1}-\ldots-\pi_{id-1}=\pi_{id}$ and $1-\sum_{k=1}^{d-1}\gamma_k = \gamma_d$, the probability density function of $\undertilde{\pi_i}=(1-\pi_{i1}-\ldots-\pi_{id})$ becomes
    \begin{align*}
        f(\pi_{i1},\ldots,\pi_{id}) & \propto \pi_{id-1}^{\alpha_0\gamma_{d-1}} \pi_{id}^{\alpha_0\gamma_d} \prod_{j=1}^{d-2} \pi_{ij}^{\alpha_0 \gamma_j}
    \intertext{Or,}
        f(\pi_{i1},\ldots,\pi_{id}) & \propto \prod_{j=1}^{d} \pi_{ij}^{\alpha_0 \gamma_j} \qquad 0<\pi_{ij}<1 \text{ and, } \sum_{j=1}^d \pi_{ij}=1 \forall i,j=1,\ldots,d
    \end{align*}
    which is the probability density function of a Dirichlet($\alpha_0\undertilde{\gamma}$) distribution.
\end{proof}

\section{Rejection Sampler for Tilted Gamma Distribution} \label{Appendix-C}
The tilted Gamma density $f_j, j=1,2,\ldots,d$ is given by 
\[
    f_j(x) \propto \frac{1}{\Gamma(x)^d} x^{\delta_j-1} e^{-B_j x} \mathbb{1}_{[x>0]}
\]
where, $d \in \mathbb{N}$ and $ \delta_j, B_j >0$. Let $\tilde{f}_j=\frac{1}{\Gamma(x)^d} x^{\delta_j-1} e^{-B_j x}$ denote the probability density function upto constants and $C_{f_j}$ the normalizing constant. 
\subsection{Rejection sampling algorithm}
 Let $g_j=\tilde{g_j}/C_{g_j}$ on $\mathbb{R}^+$ be the target density such that $\tilde{f_j}(x) \leq \tilde{g_j}(x)$ for all $x$ belonging to the support of $f_j$ for each $j=1,2,\ldots,d$. The rejection sampling algorithm is then given by:
\begin{enumerate}
    \item Draw samples $s \sim g_j$ and $u \sim$ Uniform(0,1) independently.
    \item Accept $s$ as a sample from $f_j$ if $u \leq \frac{\tilde{f_j}(s)}{\tilde{g_j}(s)}$. Else return to Step 1.
\end{enumerate}
\subsection{Construction of cover density}
\citet{das2024blocked} provides the technique for construction of cover density of the tilted gamma distribution and the procedure of sampling from it. The procedure is discussed here briefly.

When $B_j>0$, since $x>0$, $x^{\delta_j-1}, e^{-B_j x}$ is decreasing and continuous in $x$, while $\Gamma(x)$ is strictly log-convex with its minimum value lying in the interval $(1.46,1.47)$. So the mode of the density $f_j$ lies in the interval $(0,1.5)$. When $B_j<0$, the mode lies in the interval $(0, e^{1-B_j/d})$ which has been shown in the above mentioned paper. Let
\begin{align*}
    \tilde{h_j}(x) & = \log \tilde{f_j}(x) = -d\log \Gamma(x) + (\delta_j-1) \log(x) - B_jx,\\
    \tilde{h_j}'(x) & = -d \psi(x) + \frac{\delta_j-1}{x} - B_j
\end{align*}
where, $\psi(x)$ is the derivative of $\Gamma(x)$ with respect to x.
For $N \in \mathbb{N}, 2N+2$ knot points denoted by $0<m_{j,1}<\dots<m_{j,N+1}<\dots<m_{j,2N+2}<\infty$ are taken in the following manner:
\begin{enumerate}
    \item Central knot $m_{j,N+1}$ at the mode of $f_j$, obtained by solving $\tilde{h_j}'(m_{j,N+1})=0$
    \item Last knot $m_{j,2N+2} = (m_{j,N+1}+1.5)\mathbb{1}_{[B_j>0]} + e^{1-B_j/d} \mathbb{1}_{[B_j<0]}$
    \item Set the first and last but one knots as $m_{j,1} = m_{j,N+1}/2$ and $m_{j,2N+1} = (m_{j,N+1} + m_{j,2N+2})/2$.
    \item The remaining $2N-2$ knots are placed in such a way that $N_1$ knots are equidistant points between first and central knot and the rest $N-1$ are equidistant points between the central and last but one knot.
\end{enumerate}
Equation of the tangent line of $\tilde{h}_j$ at the point $m_{j,k}$ is given by
$$\nu_{j,k}(x) = a_{j,k} + \lambda_{j,k}(x - m_{j,k}),\quad k = 1,\ldots, 2N + 2$$
where $a_{j,k} = \tilde{h}_k(m_{j,k})$ and $\lambda_{j,k} = \tilde{h}_k^{'}(m_{j,k})$. Clearly, $\lambda_{j,k} > 0$ for $i \leq N, \lambda_{j,N+1}=0$ and $\lambda_{j,k} < 0$ for $k \geq N+2$. Points of intersection of the tangent lines, $\nu_{j,k}$ and $\nu_{j,k+1}$ is given by
$$q_{j,k} = \frac{a_{j,k+1} - a_{j,k} + m_{j,k}\lambda_{j,k} - m_{j,k+1}\lambda_{j,k+1}}{\lambda_{j,k}-\lambda_{j,k+1}}, \quad k = 1, \ldots,2N+1$$

Let $q_{j,0}=0, q_{j,2N+2}=\infty$ and $\nu_j(x):= \sum_{k=1}^{2N+2}\nu_{j,k}(x)\mathbb{1}_{\{x \in [q_{j,i-1},q_{k,i})\}}$. Since $\tilde{h}_j$ is concave in nature and concave functions lie below their tangent lines, we have $\tilde{f}_j(x)\leq e^{\nu_j(x)}$, which provides a piece-wise upper bound $\tilde{g}_j$ for $\tilde{f}_j$,
$$\tilde{f}_j(x) \leq \tilde{g}_j(x):= \sum_{k=1}^{2N+2}\tilde{g}_{j,k}(x)$$
where $\tilde{g}_{j,k}(x)=e^{\nu_{j,k}(x)} = e^{a_{j,k}+\lambda_{j,k}(x - m_{j,k})}\mathbb{1}_{\{x \in [q_{j,k-1},q_{j,k})\}}, k = 1, \ldots,2N+2$. The final cover density $g_j= \tilde{g_j}/C_{g_j}$ has the normalizing constant $C_{g_j} = \int_0^\infty \tilde{g}_j(x)dx = \sum_{k=1}^{2N+2}C_{g_{j,k+1}}$, where
$$C_{g_{j,k}} = \int_{q_{j,k-1}}^{q_{j,k}}\tilde{g}_{j,k}(x)dx = \begin{cases}
    e^{a_{j,N+1}(q_{j,N+1}-q_{j,N})} \text{,} \hfill i = N+1\\
    \lambda_{j,k}^{-1}e^{a_{j,k} - m_{j,k}\lambda_{j,k}}(e^{q_{j,k}\lambda_{j,k}} - e^{q_{j,k-1}\lambda_{j,k}}) \text{,} \hfill i \neq N+1
\end{cases}$$
The cover density $g_j$ can then be written as a mixture of $2N+2$ densities, $g_{j,k} =,$
\begin{align*}
    &g_j(x) = \sum_{k=1}^{2N+2}\frac{C_{g_{j,k}}}{C_{g_j}}g_{j,k}(x),\\
    &g_{j,k}(x) = \frac{\tilde{g}_{j,k}(x)}{C_{g_{j,k}}} = \begin{cases}
    (q_{j,N+1}-q_{j,N})^{-1}\mathbb{1}_{\{x \in [q_{j,N},q_{j,N+1})\}} \text{ , } \qquad k = N+1\\
   (e^{q_{j,k}\lambda_{j,k}} - e^{q_{j,k-1}\lambda_{j,k}})^{-1}\lambda_{j,k}e^{\lambda_{j,k}x}\mathbb{1}_{\{x \in [q_{j,k-1},q_{j,k})\}} \text{ , } \qquad k \neq N+1
\end{cases}
\end{align*}
\subsection{Sampling from the cover density}
To sample from the cover density, consider the weights $w_{j,k} = C_{gj,k}/C_{gj}, k = 1,\dots, 2N+2$ and $w_{j,0}=0$. To obtain a sample $s$ from $g_j$:
\begin{enumerate}
    \item Generate a random variable $u \sim \text{Uniform}(0,1)$
    \item If $u \in \left[\sum_{l=0}^{k-1} w_{k,l}, \sum_{l=0}^{k} w_{k,l} \right)$, then draw $s \sim g_{j,k}, \qquad k=1,\dots,2N+2$
\end{enumerate}
The inverse cdf method is used to get samples from $g_{j,k},k=1,\ldots,2N+2$. Since $m_{j,N+1}$ is the mode of $f_j$, $\lambda_{j,N+1}=0$ and $g_{j,N+1}$ is a $U(q_{j,N},q_{j,N+1})$ density which can be directly sampled from. To get a sample $s$ from $g_{j,k},k\neq N+1$, which are exponential densities, the inverse cdf sampler proceeds as follows:
\begin{enumerate}
    \item Draw $u \sim  \text{Uniform}(0,1)$.
    \item Set $s = G_{j,k}^{-1}(u)$, where the cdf of $g_{j,k}$ and its corresponding inverse are given by
    \begin{align*}
        G_{j,k}(x)&=\frac{e^{\lambda_{j,k}x}-e^{\lambda_{j,k}q_{j,k-1}}}{e^{\lambda_{j,k}q_{j,k}}-e^{\lambda_{j,k}q_{j,k-1}}}\mathbb{1}_{\{x \in [q_{j,k-1},q_{j,k})\}}\\
        G_{j,k}^{-1}(x) & = \lambda_{j,k}^{-1}\operatorname{log}\{ue^{\lambda_{j,k}q_{j,k}} + (1-u)e^{\lambda_{j,k}q_{j,k-1}}\}.
    \end{align*}
\end{enumerate}
\end{appendices}

\end{document}